# Instruction Adherence in Coding Agent Configuration Files

*A Factorial Study of Four File-Structure Variables*

Damon McMillan
*HxAI Australia*

**Abstract**

Frontier coding agents read configuration files (CLAUDE.md, AGENTS.md, Cursor Rules) at session start and are expected to follow the conventions inside them. Practitioners assume that structural choices (how big, where the instruction sits, whether to split across multiple files, whether to permit contradictions in adjacent files) measurably affect adherence. We report a systematic factorial study across four manipulated variables (file size 25 to 500 lines, instruction position, file architecture, conflicting-instruction presence), measuring compliance with a trivial target annotation across 1,650 Claude Code CLI sessions (16,050 function-level observations) on two TypeScript codebases, three frontier models (primarily Sonnet 4.6, with Opus 4.6 as a CLI-matched cross-model check and Opus 4.7 reported descriptively under a CLI-version confound), and five coding tasks. We use mixed-effects models with a Bayesian companion.

None of the four structural variables or three two-way interactions in the design produces a detectable contrast after multiple-testing correction. The size and conflict nulls are supported by affirmative-null Bayes factors ($BF_{10}$ between 0.05 and 0.10); the position and architecture nulls are failures to reject without Bayes-factor support.

The largest effect we measured is within-session: each additional function the agent generates within a session is associated with approximately 5.6% lower odds of compliance per generation step (OR = 0.944) within the session-length range we tested, though the relationship is non-monotonic rather than a constant per-step effect. This effect is detectable on two of the three multi-function coding tasks tested and reproduces in the same direction on a second TypeScript codebase and on Opus 4.6 at the matched configuration. This finding was identified during analysis rather than pre-specified. Within the conditions we tested, the file-structure variables we manipulated did not produce detectable contrasts in compliance. We find instead that compliance varies systematically between coding tasks and across each session's sequence of generated functions, with the shape of the within-session trend depending on the task.

## 1 Introduction

### 1.1 The rise of agent configuration files

Over the last two years, frontier coding agents have converged on the same convention: the repository ships with a plain-text configuration file (CLAUDE.md for Anthropic's Claude Code, AGENTS.md for the cross-tool Linux Foundation specification, sometimes both) which the agent reads on each session as a behavioural guide [1], [2]. Recent surveys catalogue over 2,300 agent context files across 1,925 repositories [3], and the AGENTS.md specification has uptake in 40,000-plus open-source projects [2]. Practitioners use these files to pin conventions, architectural decisions, and non-negotiable rules expected to apply across every agent invocation. The author maintains the file over time and trusts the agent to honour it on every session; the file carries no schema and no built-in confirmation that its rules have been followed, so without supporting tooling (agent hooks, guardrails, linters, CI checks) a missed instruction may go unnoticed.

### 1.2 The evidence gap

Despite how widely configuration files are used, the empirical literature on whether agents follow the instructions inside them is small. Existing studies characterise the files themselves [3], [4] or measure aggregate task success [5], and practitioner advice [6], [7], [8] is largely untested in controlled comparison. The present study is designed to vary one configuration feature at a time and measure adherence directly.

### 1.3 Research questions

Five research questions guide the work. The first four ask whether each of four manipulated features of the configuration file affects the rate at which the target instruction is followed. The fifth asks how compliance changes within a session as the agent generates more code; we identified this question during analysis after observing systematic patterns in the per-run data, and treat it as a follow-on rather than a primary research question (Table 1).

**Table 1: Research questions.**

| RQ | Research question |
|---|---|
| RQ1 | How does the size of the configuration file affect instruction compliance? |
| RQ2 | How does the position of the target instruction within the configuration file affect compliance? |
| RQ3 | How does the number of configuration files affect compliance? |
| RQ4 | Does the presence of an instruction that contradicts the target reduce compliance? |
| RQ5 | How does compliance change as the agent generates more code within a session? *(identified during analysis)* |

### 1.4 Contributions

We report a controlled experiment of 1,650 Claude Code CLI sessions that varies four configuration-file features in isolation (file size, instruction position, file architecture, and conflicting-instruction presence; see §Methodology, Table 2 for levels) on two TypeScript codebases and three frontier models (primarily Sonnet 4.6, with Opus 4.6 as a CLI-matched cross-model check and Opus 4.7 reported descriptively under a CLI-version confound). Compliance is measured as emission of a binary annotation (// @tracked) detected by AST parsing of the post-session repository state; the four no-configuration baseline cells (three on ixartz across all three models, one on Umami) show zero spontaneous emission of the annotation across 1,669 function-level observations (sentinels excluded), providing a clean zero baseline.

The study makes three contributions.

**Structural-variable nulls.** None of the four manipulated features of the configuration file produces a detectable contrast within the tested ranges. The size and conflict nulls are supported by affirmative-null Bayes factors ($BF_{10}$ between 0.05 and 0.10); the position and architecture nulls are failures to reject without Bayes-factor support.

**Within-session compliance attenuation.** Each additional function the agent generates within a session is associated with approximately 5.6% lower odds of compliance per generation step (OR = 0.944), though the relationship is non-monotonic rather than a constant per-step effect. The within-session shape varies substantially by task, and a substantial share of the attenuation concentrates in the first three to four generated functions (median first omission at generation position 4) rather than accumulating gradually. The pattern reproduces in direction on a second TypeScript codebase and on Opus 4.6 at the matched configuration. This finding was identified during analysis rather than pre-specified as a research question; we treat it as a follow-on investigation.

**Task identity matters more than file structure.** Task identity is a stronger predictor of compliance than any structural variable we manipulated. The largest within-codebase contrast across our task set is a 26.2-percentage-point gap between two specific tasks (a refactor of an existing layout component and a greenfield analytics-dashboard build, with similar average function counts per run). We report this 26.2-percentage-point contrast as a single-contrast observation between two specific tasks rather than a modify-vs-write generalisation.

## 2 Related Work

Our study sits at the intersection of three literatures: positional and length effects in language-model attention, empirical studies of agent configuration files, and the broader instruction-following benchmark tradition.

### 2.1 Positional and length effects in long contexts

The foundational observation that language models use long contexts non-uniformly is due to [9], whose multi-document question-answering experiments produced the U-shaped accuracy curve: best performance when the relevant information sits at the start or end of context, with degradation in the middle. Subsequent work has refined this finding: [10] documented degradation across 18 LLMs even when retrieval is clean; [11] tied the phenomenon to attention dilution; [12] reported that frontier models' effective context windows fall far short of advertised limits, sometimes by as much as 99%; and the NVIDIA RULER benchmark [13] found that only about half of the models claiming 32K-plus context maintain satisfactory performance at length.

The positional-attention literature is dominated by question- answering with a single retrieval target. A CLAUDE.md instruction is structurally different: it is an always-on rule, not a query answer. We test whether the U-shape transfers from within-context retrieval to within-config-file instruction position by manipulating position at five ordinal levels and file size from 25 to 500 lines.

### 2.2 Agent configuration files

The empirical study of agent configuration files is recent and small. [4] analysed 253 CLAUDE.md files from 242 repositories, and their follow-up [3] extended this to 2,303 files from 1,925 repositories. Both characterise file structure and evolution rather than test for adherence. The closest study to ours in spirit is [5] from ETH Zurich, which evaluated AGENTS.md files on a real-world Python task suite and reported that context files tend to reduce task success while inflating inference cost by over 20%; that paper documents that context files can hurt but does not isolate which features of those files account for the degradation. Practitioner guidance is similarly uncontrolled: [6] recommends "under approximately 300 lines" for CLAUDE.md, and Anthropic's long-context guidance [7] recommends placing instructions at the end of the prompt rather than near the start. None of this guidance has been validated by a controlled comparison that varies one feature at a time. Our earlier work [14] applied a controlled-comparison approach to file-level architectural decisions for structured-data accuracy on file-native agentic systems; the present study extends that approach to instruction adherence.

### 2.3 Instruction-following and its failures

Benchmarks such as LIFBench [15] and LongGenBench [16] report that adherence degrades with context length and instruction complexity. A parallel thread asks whether formal instruction hierarchies (system > user > tools) actually control model behaviour: the Instruction Hierarchy paper [17] proposed a training-time priority system, and [18] showed empirically that such hierarchies are unreliable and that models respond to societal hierarchy framings (e.g., authority and expertise) over formal priority markers. We extend this thread by testing whether explicit opposition between a CLAUDE.md instruction and an AGENTS.md counter-instruction produces a measurable compliance penalty. The informal blog literature [19] describes a compliance curve that collapses between turn 5 and turn 20 in an illustrative example; we revisit this observation by analysing compliance against within-session function position rather than turn count.

## 3 Methodology

### 3.1 Research design

We use a fractional factorial design with four independent variables (file size, instruction position, file architecture, and conflict presence) and a binary behavioural dependent variable measured at the function level. A full factorial of all levels would require 120 cells; we instead test the four main effects and three two-way interactions that correspond to the research questions in this study, a design that preserves power on the comparisons of interest while keeping the run budget tractable. Analyses identified after data collection are described as follow-on investigations where they appear.

### 3.2 Independent variables

We manipulate four features of the configuration file. Table 2 summarises the levels of each and the reference cell at which the other variables are held when the IV is varied. Conflict requires a secondary file, so C1 conditions necessarily use the A2 (or A3) architecture; the conflict-by-position sub-block estimates the conflict effect at P3 and P5.

**Table 2: Independent variables and reference cell.**

| IV | Levels | Reference (others held at) |
|---|---|---|
| File size | 25, 100, 250, 500 lines (S1 to S4) | P1 / A1 / C0 |
| Instruction position | top, ≈25%, centre, ≈75%, bottom (P1 to P5) | S3 / A1 / C0 |
| File architecture | single CLAUDE.md, + AGENTS.md, + nested per-directory files (A1 to A3) | S3 / P1 / C0 |
| Conflicting instruction | absent, present (C0, C1) | S3 / A2 / P3+P5 |

**File size (IV1).** The four levels (25, 100, 250, 500 lines) span from minimal to large within the empirical range of CLAUDE.md line counts documented in surveys of public repositories [4]. Each variant holds the target instruction constant and varies only the surrounding content. Padding content was authored to mirror the topical categories that [4] document as common in real CLAUDE.md files; it is not extracted verbatim from any specific repository.

**Instruction position (IV2).** Position is varied at a fixed 250-line CLAUDE.md, with the target instruction at line 2 (P1), line 63 (P2), line 128 (P3), line 187 (P4), or line 250 (P5). Five levels are enough to distinguish the two qualitatively different shapes reported in the positional-attention literature: monotonic primacy decline [20] and a symmetric U-shape reflecting combined primacy and recency [9], while providing power on the centre position. The design also draws on the needle-in-a-haystack tradition [21], particularly [22]'s finding that smaller needles amplify positional sensitivity, motivating the choice of a single-line target instruction. For the main-effects block, position is manipulated at S3 only, because a 25-line file does not admit a meaningful positional manipulation. The size-by-position sub-block additionally tests positions P3 and P5 at S2 and S4.

**File architecture (IV3).** All architectures hold the target instruction in the root CLAUDE.md; the manipulation affects only the surrounding instruction surface. A2 adds an AGENTS.md file containing complementary (non-conflicting) project context [2]; A3 extends A2 with two nested CLAUDE.md files in src/ and tests/ carrying directory-specific conventions. Content for AGENTS.md and the nested files was authored to reflect the structure and topical mix that [3] report as common across the 2,303 agent-context files they survey, but is not extracted verbatim from any external corpus.

Claude Code reads the root CLAUDE.md automatically on session start but does not read AGENTS.md by default. To ensure that AGENTS.md was loaded into context in the A2 and A3 conditions, our injection script prepended an @AGENTS.md import directive to the root CLAUDE.md in those conditions. The directive is the standard Claude Code mechanism for referencing additional configuration files; the target instruction is unchanged in either file by this addition. Nested CLAUDE.md files (src/CLAUDE.md, tests/CLAUDE.md) are read by Claude Code when work touches the corresponding directories.

**Conflict presence (IV4).** In C1 conditions, a secondary file carries an instruction directly contradicting the target. The verbatim text in AGENTS.md reads: "Keep function bodies clean and readable. Do not add non-documentation comments inside function bodies. Comments should only appear above functions as JSDoc, never inside the function body itself." Because the conflict lives in AGENTS.md, C1 conditions necessarily use A2 or A3 architecture. This factor tests whether instructions are honoured when they collide across files [17].

### 3.3 Dependent variable

The target instruction is a syntactically detectable marker: the agent is told to begin every new or modified function body with the comment // @tracked as its first line, justified to the agent as a requirement of an internal code-audit system. The marker has five properties that make it tractable as a dependent variable: it is binary-measurable; it never occurs by default in baseline runs; it applies to every coding task in our test suite; it produces one observation per function; and its simplicity rules out comprehension failure as an explanation for non-compliance. We treat the marker as an *easily-detectable proxy* for instruction adherence rather than a measure of adherence in general: more demanding instructions are subject to additional failure modes (comprehension, multi-step reasoning, conditional logic) that the marker does not capture, so transfer of the absolute compliance rate to other instruction classes is an open question.

Compliance is measured at the function level. After each run we compute the git diff against the pre-task baseline, extract every new or modified function body via an abstract syntax tree of the TypeScript and JSX source, and check whether the body's first statement is exactly // @tracked. The per-run Instruction Compliance Rate (ICR) is the ratio of compliant to total functions. Statistical analyses retain the function-level binary outcome rather than collapsing to a per-run rate, both to preserve statistical power and to permit within-run analyses such as the per-position attenuation curve.

### 3.4 Test repositories

The primary corpus is the ixartz/Next-js-Boilerplate, a widely starred TypeScript and Next.js scaffold pinned to a specific commit for the duration of the study. The baseline state of the repository contains no CLAUDE.md, AGENTS.md, or .claude/ directory. For cross-codebase validity we replicate a subset of conditions on **Umami**, a larger analytics platform (663 TypeScript files at the pinned commit), which exercises whether the findings generalise beyond a minimal boilerplate.

### 3.5 Condition matrix

The 24 experimental conditions are organised into four factorial sub-blocks plus a no-configuration baseline. **Main effects (11 conditions, ME-01 to ME-11)** isolate each IV in turn while holding the other three at a reference level (S3, P1, A1, C0): four file sizes, five instruction positions, three architectures, and one conflict condition. **Size x Position (4 conditions, SP-01 to SP-04)** crosses sizes S2 and S4 with the middle and bottom positions. **Size x Architecture (4 conditions, SA-01 to SA-04)** crosses sizes S2 and S4 with the two multi-file architectures. **Conflict x Position (4 conditions, CP-01 to CP-04)** crosses conflict and its absence with the middle and bottom positions under the A2 architecture. The **baseline (BL-01)** runs without any configuration files to confirm the zero-compliance floor. Each ixartz condition was executed with 50 runs. The ecological replication contributes three Umami conditions (EV-01 to EV-03) at 50 runs each. Cross-model generalisation contributes three Opus 4.6 conditions (OP-01 to OP-03) at 50 runs each, and three Opus 4.7 conditions (OP-01 to OP-03) at 50 runs each. The full per-condition matrix is given in Appendix A.

### 3.6 Coding tasks

Each ixartz run is assigned one of five coding tasks, with the design fully crossed: every task runs 10 times per ixartz condition. The tasks (Table 3) span the space of typical feature-addition work on a Next.js boilerplate. Average functions per run are computed across the ixartz Sonnet 4.6 primary pool. T4 is the only task whose primary work is modifying existing code; the others ask the agent to write new code. T1 and T2 are single-function tasks (1.0 and 1.2 functions per run on average) included as complexity-floor controls: they verify that compliance is high under low-complexity conditions and exclude the simplest-of-tasks alternative explanation for the structural-variable nulls. The within-session attenuation analysis is fitted on the multi-function tasks (T3, T4, T5) only, because T1 and T2 do not produce sufficient positional sweep within a session.

**Table 3: Coding tasks.**

| Task | Description | Complexity | Avg functions / run |
|---|---|---|---|
| T1 | New /api/health endpoint | low | 1.0 |
| T2 | Reusable Button component with variants | low-medium | 1.2 |
| T3 | NextAuth integration | medium | 16.0 |
| T4 | Sidebar refactor (modify existing layout) | medium-high | 17.5 |
| T5 | /dashboard/analytics page with loading skeletons | high | 15.0 |

Task is treated as a nuisance factor and enters the GLMM as a random effect rather than an additional IV, so that inferences generalise across tasks rather than conditioning on any one of them.

### 3.7 Procedure and environment

Each run followed an automated procedure: the test repository was reset to its pinned baseline commit, the condition's configuration files were injected into the root (and into src/ and tests/ for A3), the Claude Code CLI was invoked in headless mode against the task prompt, and the resulting diff was scored by the AST-based pipeline described above. Every run recorded its condition, task, run index, git hash, configuration-file SHA-256 hashes, exact model ID, token counts, turn count, and wall-clock duration to a per-block SQLite database conforming to a frozen schema.

The Claude Code CLI was pinned at version 2.1.92 for the Sonnet 4.6 and Opus 4.6 blocks. The Opus 4.7 block used CLI version 2.1.123, required by an API schema change for adaptive thinking that CLI 2.1.92 does not emit. Both CLI versions were pinned to isolated installations to prevent auto-update contamination. In the cross-model blocks, OP-01 shares the ME-03 configuration parameters (S3 / P1 / A1 / C0) as a clean reference cell, OP-02 is the "worst-case" stress cell that combines all three structural levers known to be harder (S3 / P3 / A2 / C1: a mid-positioned target instruction in a split-file architecture with a contradicting instruction in AGENTS.md), and OP-03 is the no-configuration baseline. The cross-model design is therefore "reference + extreme + baseline" rather than a within- Opus replication of the structural-variable manipulations on Sonnet; the OP-01 versus OP-02 contrast varies architecture, position, and conflict simultaneously and does not isolate any one of them. The CLI difference is disclosed alongside any cross-model claim. Sessions were permitted up to 100 agent turns; the ceiling was never binding (median 14, maximum 81 across the 1,650 sessions).

A subset of Opus 4.7 (run, task) trials returned a clarifying question rather than code in the single-turn --print mode used by the harness. These trials produced no functions for compliance scoring and are reported as a Code-Production Rate metric alongside ICR; we treat ICR for Opus 4.7 as conditional on the model producing code.

### 3.8 Sample size

Each condition was allocated 50 runs across all blocks. Simulation-based power analysis (R's simr package, anchored on the ixartz Sonnet 4.6 primary pool) returns at least 98% power at $n = 50$ to detect a 15-percentage-point ICR shift on the most-extreme level of each IV using the planned GLMM factor-LRT. For a 10-percentage-point shift, power is 99% (conflict, 1-df), 99% (position, 4-df), 82% (architecture, 2-df), and 70% (size, 3-df); the size IV is the least sensitive to small effects because its 3-df factor LRT is a harder test than the lower-df tests on the other IVs. Designs aiming to detect specifically 10-percentage-point differences on the size factor would benefit from $n > 50$.

All reported analyses operate on a single frozen merged dataset of 1,650 sessions (1,200 ixartz Sonnet 4.6, 150 Umami Sonnet 4.6, 150 Opus 4.6, 150 Opus 4.7) and 16,050 function-level observations after excluding 43 sentinel rows where the agent produced no code (Opus 4.7: 11 on OP-01, 7 on OP-02, 20 on OP-03; Opus 4.6: 1 on OP-01, 2 on OP-02, 2 on OP-03; Sonnet 4.6 produced code on every trial). The sentinel rows are reported separately as a Code-Production Rate metric in Results.

### 3.9 Statistical analysis plan

The primary analysis described in advance is a binomial generalised linear mixed model with function-level compliance as outcome, a logit link, and the structure

$$\text{Compliance}_{ij} \sim \text{Size}_i \times \text{Position}_i \times \text{Architecture}_i \times \text{Conflict}_i + (1 \mid \text{RunID}_j) + (1 \mid \text{TaskID}_k).$$

Random intercepts for RunID and TaskID absorb within-run correlation and task-level mean differences. Because the fractional factorial does not include every cell of the full four-way factorial, the four-way interaction is not estimable as specified; we therefore fit the main-effects model against the main-effects block, and the three two-way interactions against their respective sub-blocks, as four separate GLMMs whose union recovers the effect structure.

For the within-session attenuation analysis we operationalise function position in two ways. The primary predictor is *chronological generation order*: a 1-based linear rank over the agent's file-touching tool calls within the session, reconstructed from each session's tool-call timeline. Every function in the merged scoring pipeline is matched to its earliest emission rank (over 99% by exact (file, name, body) match; the remainder by relaxed match where the function was edited after emission). Functions in files the agent never touched in the session have no chronological position and are excluded by construction. We report chronological order as primary because it directly indexes the sequence a practitioner observes. The secondary predictor is the AST-traversal function index (alphabetical files, declaration-order traversal), reported as a sensitivity; it incorporates within-file structural position and does not isolate the within-session signal as cleanly.

All four per-IV main-effect tests, the three two-way interaction tests, and the within-session attenuation analysis are fitted as binomial GLMMs via lme4::glmer in R using the bobyqa optimiser, on the appropriate sub-blocks of the merged dataset. We report the GLMM likelihood-ratio test as the primary inferential test for each contrast and report simpler descriptive cross-checks (pairwise chi-squared with Benjamini-Hochberg correction, run-level ANOVA, fixed-effects logistic-regression LRT) alongside in Results where they aid interpretation; the GLMM is the canonical test because it preserves the function-level random structure ((1|run_id) + (1|task_id)) that the design generates.

A Bayesian companion is fitted with bambi on PyMC using four chains of 2,000 post-warmup draws each and target_accept = 0.99 (raised from the standard 0.95 because the (1|task_id) random intercept tightens posterior geometry, and an initial fit at 0.95 produced more than the prior-specified 1% divergence ceiling on the tightest prior). Priors are bambi's default autoscaled weakly-informative Normals on each fixed effect (this corresponds approximately to Normal(0, 1) on standardised size and Normal(0, 2) on the binary conflict predictor, depending on each predictor's standard deviation; the sensitivity sweep below tests $\sigma$ values spanning these defaults). Bayes factors for point-null hypotheses use the Savage-Dickey density ratio with the prior density at zero estimated from 8,000 prior samples rather than a parametric approximation. Bayes factors are interpreted on the Jeffreys scale. Robustness of the headline Bayes factor claims to prior choice is checked via a prior-sensitivity sweep on Normal(0, $\sigma$) priors for $\sigma \in \{0.5, 1.0, 2.5, 5.0\}$, reported in Results §"Bayesian robustness".

Frequentist tests use $\alpha = 0.05$ two-tailed. Post-hoc pairwise comparisons are corrected by the Benjamini-Hochberg procedure at FDR $q = 0.05$, applied per analysis with raw p-values stored alongside the adjusted values. Planned GLMM contrasts following the research questions are not corrected separately. Effect sizes are reported as Cohen's $h$ for proportions, $\omega^2$ for ANOVA-level comparisons, and odds ratios with Wald 95% confidence intervals for GLMM coefficients; Bayesian intervals are 95% credible intervals. Wilson score intervals are used for single-proportion CIs where exact rates approach 0 or 1.

### 3.10 Research questions

**RQ1 (file size).** How does the size of the configuration file affect instruction compliance? We test four sizes spanning a wide range within the empirical distribution of CLAUDE.md files (S1 = 25 lines through S4 = 500 lines).

**RQ2 (instruction position).** How does the position of the target instruction within the configuration file affect compliance? We test five ordinal positions from P1 (top) to P5 (bottom).

**RQ3 (file architecture).** How does the number of configuration files affect compliance? We test three levels: a single CLAUDE.md (A1), CLAUDE.md plus AGENTS.md (A2), and CLAUDE.md plus AGENTS.md plus two nested CLAUDE.md files (A3).

**RQ4 (inter-file conflict).** Does the presence of an instruction that contradicts the target reduce compliance? We test C0 (no conflict) and C1 (a conflicting instruction in a secondary file). The conflict effect is estimated at the middle (P3) and bottom (P5) positions only; the conflict-at-top cell is not in the design.

**RQ5 (within-session change in compliance).** Sessions produce a variable number of functions (1 to approximately 25 per run); does compliance change systematically with the ordinal index of a function within the session?

We also test three two-way interactions of the four manipulated variables (size x position, size x architecture, conflict x position) and report those alongside the corresponding RQs.

## 4 Results

The full study comprises 1,650 sessions and 16,050 function-level observations (after excluding 43 sentinel rows where the agent produced no code) across two codebases and three frontier models. The results that follow are stratified by pool. The structural- variable analyses are estimated on the *ixartz Sonnet 4.6 primary pool* (1,150 non-baseline runs, 11,637 functions); the no- configuration baseline (BL-01, 524 functions), the Umami ecological block (3 conditions, 1,109 functions), the Opus 4.6 cross-model block (1,703 functions after sentinel exclusion), and the Opus 4.7 cross-model block (1,077 functions after sentinel exclusion) are reported in their own subsections, summing to the full 16,050. Throughout, $n$ values refer to the ixartz Sonnet 4.6 primary pool unless stated otherwise. The 43 sentinel rows are reported separately as Code-Production Rate.

### 4.1 Structural variables show no detectable effect within tested conditions

The design includes four manipulated independent variables (size, position, architecture, conflict) and three two-way interactions among them. None of the seven contrasts is detectable at $\alpha = 0.05$ after Benjamini-Hochberg correction; the Bayesian companion returns affirmative-null evidence for size and conflict. We report each in turn.

**File size.** ICR across the four sizes (S1 = 25, S2 = 100, S3 = 250, S4 = 500 lines), all at P1, A1, C0, was 60.0%, 65.2%, 67.7% and 64.0% respectively. The planned GLMM main-effect test (binomial GLMM with random intercepts for run and task) is $\chi^2 = 5.16$, $df = 3$, $p = 0.16$. As descriptive cross-checks, all six pairwise chi-squared tests are non-significant after BH-FDR correction (smallest BH $p = 0.077$); the run-level one-way ANOVA is $F(3{,}196) = 0.49$, $p = 0.687$, $\omega^2 = 0.000$; the linear trend across size is $p = 0.625$.

**Instruction position.** ICR across the five positions (P1 line 2, P2 line 63, P3 line 128, P4 line 187, P5 line 250), all at S3, A1, C0, was 67.7%, 63.2%, 64.0%, 64.0% and 61.8% respectively. The planned GLMM main-effect test is $\chi^2 = 1.47$, $df = 4$, $p = 0.83$. Orthogonal-polynomial decomposition under the same GLMM is also null at every order: linear $\chi^2(1) = 0.68$, $p = 0.41$; quadratic $\chi^2(1) = 0.01$, $p = 0.94$; cubic $\chi^2(1) = 0.73$, $p = 0.39$; quartic $\chi^2(1) = 0.05$, $p = 0.83$. The largest contrast (P1 vs P5 at $-5.9$ pp, raw $p = 0.055$) does not survive BH-FDR correction over the ten pairwise comparisons (BH $p = 0.554$); the descriptive pattern shows weak primacy at P1 followed by a small monotonic drift, but no individual contrast reaches significance, and no polynomial shape component is detectable.

**File architecture.** ICR across the three architectures (A1 single CLAUDE.md, A2 CLAUDE.md plus AGENTS.md, A3 A2 plus two nested CLAUDE.md files), all at S3, P1, C0, was 67.7%, 68.2% and 61.7% respectively. The planned GLMM main-effect test is $\chi^2 = 3.29$, $df = 2$, $p = 0.19$. A2 is statistically indistinguishable from A1 ($+0.5$ pp, $p = 0.918$); the suggestive A3 dip ($-6.0$ pp vs A1, raw $p = 0.053$) does not survive BH correction (smallest BH $p = 0.079$). The run-level ANOVA is $F(2{,}147) = 1.62$, $p = 0.20$, $\omega^2 = 0.008$.

**Conflicting instruction.** The cleanest test is the four-cell factorial in the conflict-by-position sub-block where both P3 and P5 appear with and without conflict, all at S3 and A2 (CLAUDE.md plus AGENTS.md). The planned GLMM main-effect test is $\chi^2 = 0.10$, $df = 1$, $p = 0.76$. Descriptively, the pooled C0 vs C1 contrast on this balanced design is $+0.33$ pp (C0 = 63.7%, C1 = 64.1%; $\chi^2 = 0.012$, $p = 0.912$, Cohen's $h = -0.007$, $n = 2{,}044$); none of the pairwise contrasts survives correction.

**Across the four manipulated structural variables, no contrast is detectable within the tested ranges.**

**Two-way interactions.** All three two-way interactions return null GLMM likelihood-ratio tests: size × position ($\chi^2 = 0.42$, $df = 2$, $p = 0.81$), size × architecture ($\chi^2 = 0.81$, $df = 2$, $p = 0.67$), and conflict × position ($\chi^2 = 2.27$, $df = 1$, $p = 0.13$).

The four nulls divide into two evidentiary classes. The size and conflict nulls are supported by the Bayesian companion at $BF_{10} = 0.096$ and $BF_{10} = 0.053$ respectively under the planned random-effects structure (data favour the null over a unit-effect alternative by approximately ten to twenty on the Jeffreys scale; affirmative-null evidence within the tested ranges). The position and architecture nulls do not have Bayes-factor support; the design's 95% Wald CI half-width on the largest pairwise contrasts (±5.9 pp) rules out structural effects above approximately 6 percentage points on these two variables but cannot rule out smaller ones.

**Figure 2: Independent variable effects.**

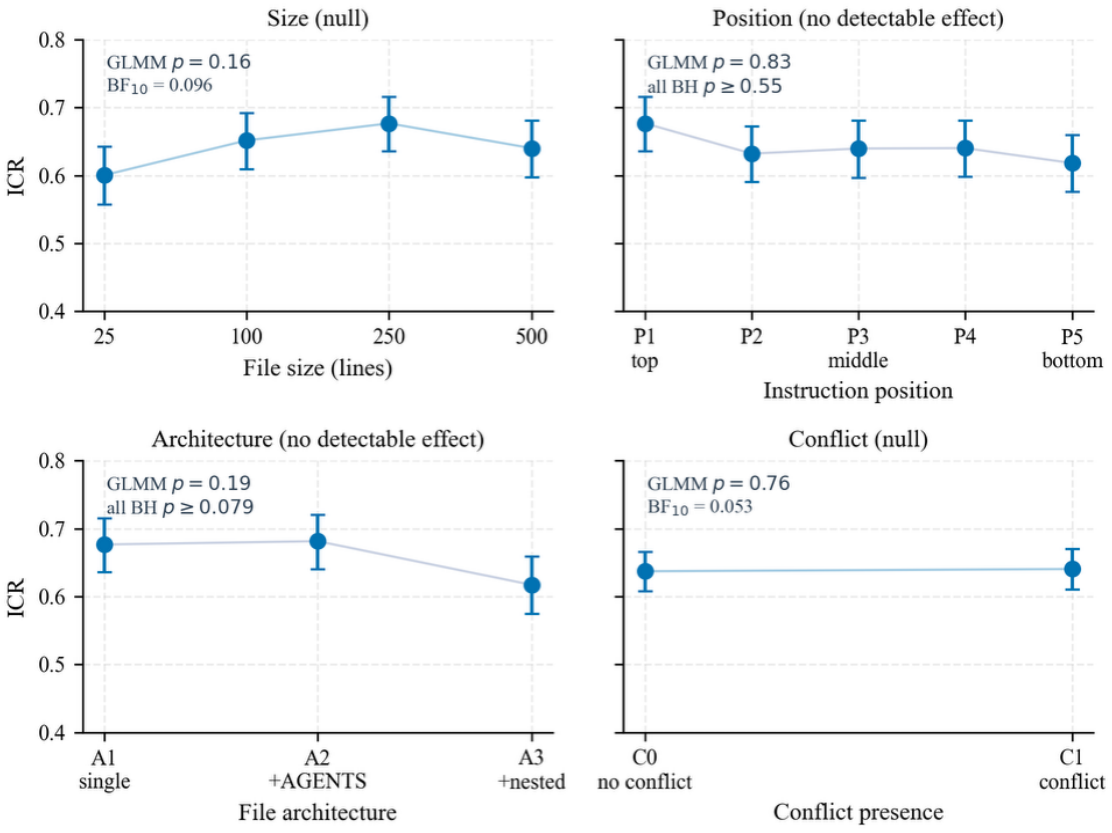

**Independent variable effects. Function-level ICR with Wilson 95% confidence intervals at each level of the four manipulated variables. Size (top-left), position (top-right), architecture (bottom-left) and conflict (bottom-right) all show no contrast surviving BH correction. Reference cell ME-03 (S3 / P1 / A1 / C0) is marked.**

### 4.2 Task type as a single-contrast observation

Across the five tasks held constant in their prompts, function- level ICR varied from 45.1% on T4 to 84.4% on T2 (Table 4).

**Table 4: Per-task instruction compliance.**

| Task | Description | Type | Runs | Functions | ICR | Avg/run |
|---|---|---|---|---|---|---|
| T1 | Health API endpoint | new code (low complexity) | 230 | 230 | 82.6% | 1.0 |
| T2 | Button component | new code (low-medium) | 230 | 269 | 84.4% | 1.2 |
| T3 | NextAuth integration | new system (medium) | 230 | 3,671 | 72.7% | 16.0 |
| T4 | Sidebar refactor | modify existing (medium-high) | 230 | 4,028 | 45.1% | 17.5 |
| T5 | Analytics dashboard | new code (high complexity) | 230 | 3,439 | 71.3% | 15.0 |

The most directly comparable pair is T4 (modify existing code, 17.5 functions per run on average) and T5 (write new high- complexity code, 15.0 per run on average): both span the full positional range and produce substantial multi-function output. T4 sits 26.2 percentage points below T5 (45.1% vs 71.3%), the largest within-codebase contrast in the data and larger than the separation between any two structural conditions. The unified-GLMM task fixed effect (LRT $\chi^2 = 355.16$, $p = 1.4 \times 10^{-75}$) confirms that task identity is a substantial source of compliance variability.

T4 is the only task in our design that asks the agent to modify pre-existing code, so the modify-vs-write read on this gap rests on a single between-task contrast and warrants follow-up with multiple modify-existing and write-new tasks of matched complexity. Whatever the underlying driver, task identity is a stronger predictor of compliance than any structural variable we manipulated: the task-level random-intercept variance is $\sigma^2_{\text{task}} = 0.310$ and the unified-GLMM fixed-effect odds ratio is T4 OR = 0.243 versus the T1 reference.

### 4.3 Baseline confirmation

The no-configuration baseline (BL-01) produced 0 compliant functions out of 524 across 50 runs (ICR = 0.0%; Wilson 95% upper bound 0.73%). The 67.7-percentage-point gap against the matched with-configuration condition ME-03 ($n = 529$, ICR = 67.7%) is large: $\chi^2 = 534.27$, $p = 3.33 \times 10^{-118}$, Cohen's $|h| = 1.93$. No model in our trials emitted // @tracked without the configuration file. The compliance we observe in every other condition is a response to the test instruction, rather than a coincidental annotation; the design is therefore measuring instruction-driven behaviour. The Wilson upper bound (0.73%) bounds the maximum plausible spontaneous-emission rate at this sample size.

### 4.4 Cross-codebase consistency

Three Umami conditions ($n = 50$ each on a 663-file analytics application) test whether the headline pattern reproduces on a second TypeScript codebase. EV-02 (no configuration) replicates the zero baseline (0 of 338 compliant; Wilson upper 1.1%). EV-01 (matched A1/S3/P1/C0 configuration) yields ICR = 56.7%, which is 11.0 percentage points below the matched ixartz reference ME-03 ($\chi^2 = 11.11$, $p = 0.0009$, $|h| = 0.228$, small). The within-Umami EV-01 (A1) vs EV-03 contrast is +0.6 pp ($\chi^2 = 0.01$, $p = 0.93$, $|h| = 0.011$, negligible); EV-03 combines A2 architecture with C1 conflict, so this contrast is not a clean architecture comparison and we do not treat it as independent confirmation of the ixartz architecture null. Absolute compliance shifts down on the larger codebase.

The within-session attenuation slope reproduces in the same direction across the three CLI-matched cells (ixartz x Sonnet 4.6, ixartz x Opus 4.6, and Umami x Sonnet 4.6, all on Claude Code CLI 2.1.92). A binomial GLMM with a cell × generation_position interaction returns a likelihood-ratio test for the interaction of $\chi^2 =$ $2.23$, $df = 2$, $p = 0.327$ on the CLI-matched pool (1,474 function-level observations from 149 runs); the per-cell slopes (logodds per generation step) are $-0.040$ for ixartz Sonnet 4.6 (OR = 0.960, 95% CI [0.928, 0.994]), $-0.029$ for ixartz Opus 4.6 (OR = 0.971, 95% CI [0.945, 0.999]) and $-0.068$ for Umami Sonnet 4.6 (OR = 0.934, 95% CI [0.894, 0.976]). All three slopes are negative and individually detectable; the LRT does not detect a difference in slope across the three cells, which is consistent with a common slope but does not constitute affirmative evidence for slope equivalence at this sample size. A two-cell ixartz-only sensitivity (drop Umami) returns $\chi^2 = 0.15$, $df = 1$, $p = 0.70$. The Opus 4.7 cells run on a different CLI version and are not part of this comparison; see Limitations.

**Figure 3: Within-session attenuation slope reproduces in the same direction across the three CLI-matched cells.**

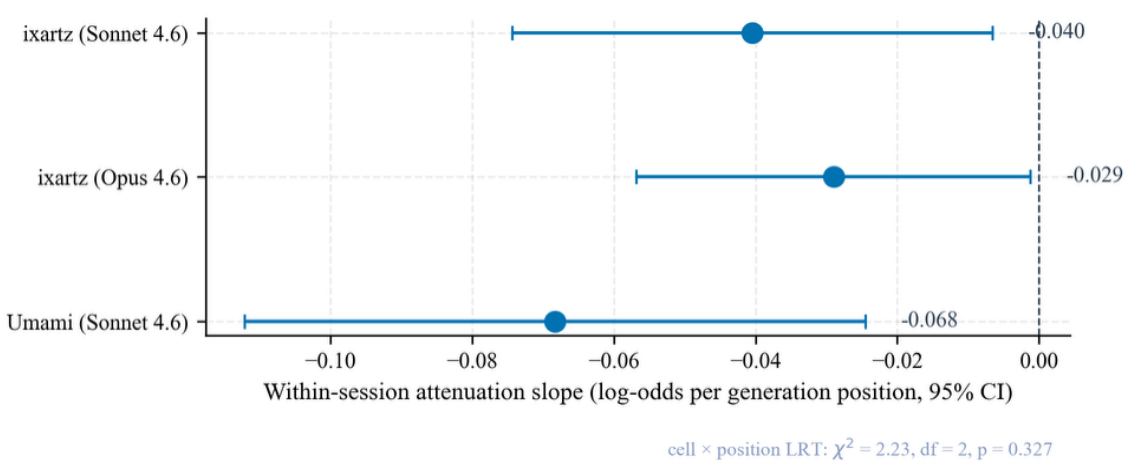

**Forest plot of within-session attenuation slopes (log-odds per generation position) for the three CLI-matched cells, with 95% Wald confidence intervals shown as horizontal whiskers. All three slopes are negative and the confidence intervals overlap; the formal cell × generation_position likelihood-ratio test on the pooled GLMM returns $\chi^2 = 2.23$, $df = 2$, $p = 0.327$ (no detectable difference, consistent with but not affirmative evidence for slope equivalence). The Opus 4.7 row is omitted because it ran on Claude Code CLI 2.1.123 with adaptive thinking enabled by default, a confound that prevents direct comparison with the CLI-matched cells.**

### 4.5 Cross-model observations

We tested two additional frontier models. The OP-01 reference cell uses the matched A1 / S3 / P1 / C0 configuration; OP-02 is a separately-framed worst-case stress cell (see Methodology §"Procedure and environment"). **Opus 4.6** ($n = 49$ valid runs; OP-01) produced ICR = 54.9%, 12.8 percentage points below Sonnet 4.6 at the same configuration ($\chi^2 = 18.70$, $p < 0.0001$, $|h| = 0.264$, small). The Opus 4.6 baseline (OP-03) replicated at 0 of 544 compliant (Wilson upper 0.70%). The OP-02 cell on Opus 4.6 (worst-case A2 / P3 / C1) produced ICR = 57.7%, $+2.8$ percentage points above OP-01 ($p = 0.36$, not significant); the contrast varies three IVs simultaneously and is reported descriptively.

**Opus 4.7** behaves differently. Of 150 (run, task) trials, 38 (25.3%) returned a clarifying question rather than code in the single-turn --print harness used by the orchestrator. We report this as a Code-Production Rate metric alongside ICR. With a CLAUDE.md present, Opus 4.7 produced code on 82 of 100 trials (CPR = 82.0%); without one, on 30 of 50 trials (CPR = 60.0%, ICR = 0% on the 263 functions produced in those 30 trials, matching the zero-baseline pattern across all four no-configuration cells). The 38 non-production cells concentrate on two of the three multi-function tasks (T3 and T4 account for 36 of 38; T5 is unaffected). On the trials where the model did produce code, ICR was 82.0% on OP-01 (clean reference) and 81.2% on OP-02 (worst-case A2 / P3 / C1 combination), a difference of $-0.8$ percentage points in the opposite direction to Opus 4.6 and again not significant. A combined GLMM on the three configuration-matched cells (Sonnet 4.6 ME-03 + Opus 4.6 OP-01 + Opus 4.7 OP-01) finds no detectable model × position interaction (LRT $\chi^2 = 0.25$, $df = 2$, $p = 0.88$).

The Opus 4.7 numbers cannot be cleanly attributed to the model: CLI 2.1.123 (required for Opus 4.7) enables adaptive thinking by default, confounding model effect with CLI/thinking configuration. A separate model × position interaction GLMM that includes Opus 4.7 in the broader chronological pool returns a detectable interaction ($\chi^2 = 16.72$, $df = 2$, $p = 0.0002$), but this detection is driven by Opus 4.7's CLI/thinking-confounded steeper slope and is not informative as a model-level finding; see Limitations.

### 4.6 Within-session compliance attenuation

This subsection reports the strongest exploratory signal in our data. The within-session question was identified during analysis from descriptive inspection of per-run patterns and was *not* pre-specified as a research question (see the *Research questions* subsection of the Introduction); we therefore report the within- session attenuation as a follow-on investigation that warrants independent replication before being treated as a robust effect.

A binomial GLMM fit to the primary pool with random intercepts for run and task and a single fixed effect for generation position estimates a slope of $-0.0578$ log-odds per generation step (SE = 0.0039, $z = -14.82$, $p = 1.08 \times 10^{-49}$). Each additional function the agent generates within a session is associated with 5.6% lower odds of compliance (OR = 0.944, 95% Wald CI [0.937, 0.951]). It is the largest effect we measured. The pooled estimate averages over substantial per-task variation reported below; the 5.6% headline figure is the primary-pool average, not a per-task constant.

Adding a fixed effect for task identity barely shifts the slope ($-0.0578 \rightarrow -0.0575$) and yields a highly significant likelihood-ratio test against the position-only model (LRT $\chi^2 = 355.16$, $df = 4$, $p = 1.4 \times 10^{-75}$); AIC drops from 14,046.6 to 14,027.1. The attenuation holds after controlling for task identity, but the slope also varies substantially across tasks. An attenuation-by-task interaction model fitted on the multi-function tasks T3, T4 and T5 is highly significant (LRT $\chi^2 = 293.17$, $df = 2$, $p = 2.2 \times 10^{-64}$). Per-task slopes (T3 as reference): T3 $+0.005$ log-odds per generation step (OR $= 1.005$, $p = 0.44$, not significant at this sample size); T4 $-0.046$ (OR $= 0.955$, $p = 1.0 \times 10^{-8}$); T5 $-0.185$ (OR $= 0.831$, $p = 1.4 \times 10^{-60}$). T1 and T2 are singleton tasks (1.0 to 1.2 functions per run on average) and are excluded from the interaction fit; positional analysis is not viable for them. The pooled GLMM remains the headline test; the per-task slopes describe heterogeneity within the pool rather than independent claims.

**Figure 1: Within-session compliance attenuation.**

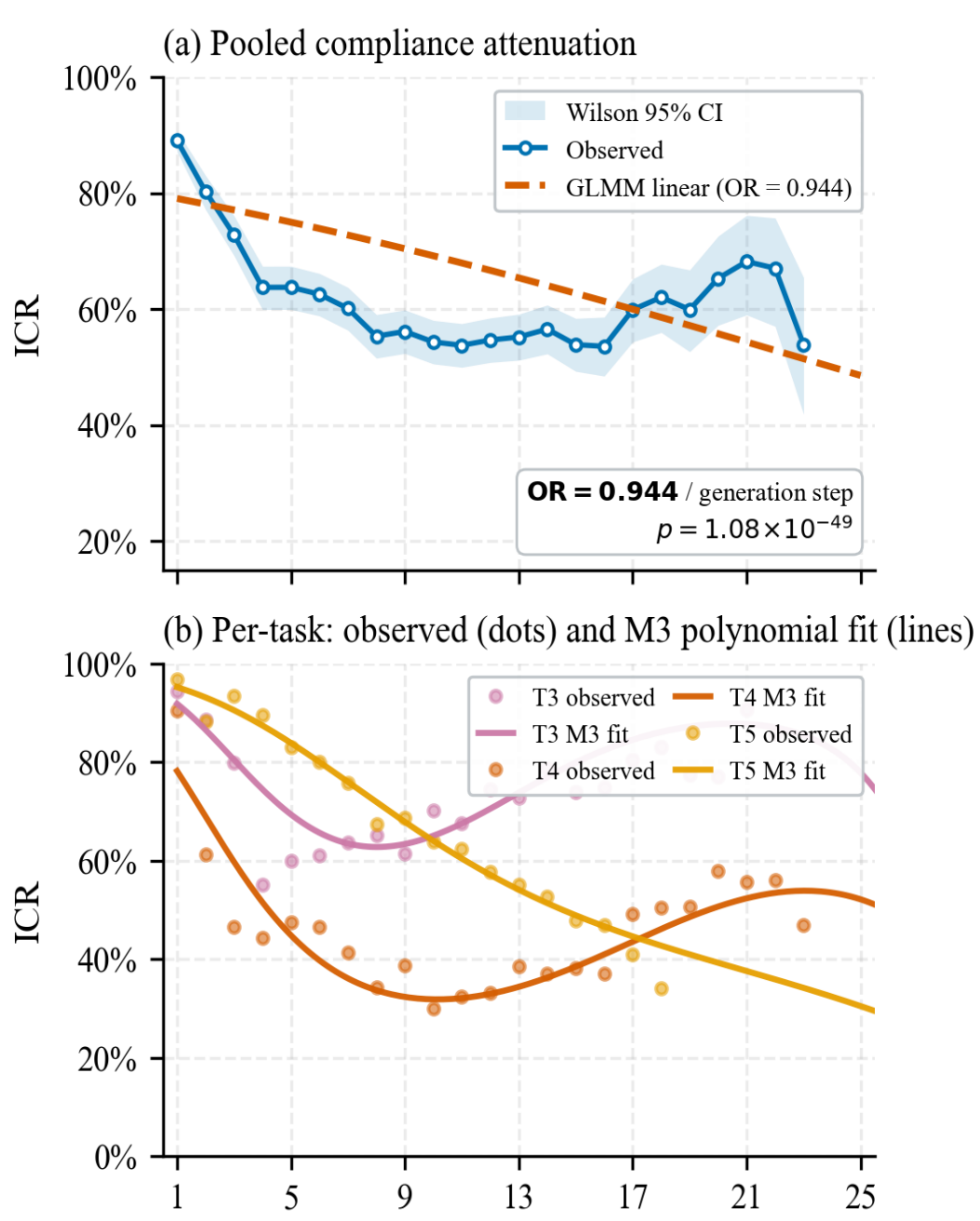

**Within-session compliance attenuation on generation order, plotted across positions 1 to 25 (the full well-powered range plus a small slice of the noisy tail at high positions where sample sizes thin). Panel (a): pooled instruction compliance rate (ICR) by generation position within a session, with the Wilson 95% confidence ribbon, observed rates (circles, n ≥ 50 per position), and the fitted GLMM logistic attenuation curve (dashed) at OR = 0.944 per generation step ($p = 1.08 \times 10^{-49}$). Panel (b): per-task observed dots (n ≥ 30 per position) with the M3 polynomial fit (linear + quadratic + cubic, fitted by task) overlaid as a solid line per task. The polynomial fits show that T3 has a flat linear average masking detectable quadratic and cubic shape (early decline followed by recovery toward the well-powered tail), T4 has shape components across all three orders, and T5 is dominated by the linear attenuation component (cubic component not detectable for T5).**

**Sensitivity: higher-order chronological shape.** Nested likelihood-ratio tests on orthogonal-polynomial contrasts of chronological generation position show the pooled slope averages over a non-monotonic shape: linear, quadratic and cubic components are all detectable on the primary pool ($\chi^2 = 216$, 313, and 99; all $p < 10^{-22}$), and all three remain detectable when restricted to the well-powered range (gen_pos $\leq 15$, $n \geq 455$ per position), ruling out a high-position-tail artefact. Per-task profiles differ: T5 is dominated by linear, T4 has detectable shape across all three orders, and T3 has a non-detectable linear masking detectable quadratic and cubic components (consistent with an early decline followed by recovery toward the well-powered tail). The pooled OR = 0.944 should therefore be read as the average per-step effect across the full within-session range; the actual per-step effect varies with generation position and task.

**Compositional shift in the late-session recovery.** The marginal recovery at high generation positions reflects a compositional shift toward categories with stable high compliance (page and layout) rather than the agent re-engaging with the configuration. We discretise generation position into three buckets (early decline: gen_pos 1-3; plateau: 4-15; recovery: 16-23) and report within-category trends across them. T4 and T5 component files show consistent compliance regression across these buckets (T4: 0.66 → 0.35 → 0.24, $\chi^2 = 238$, $p < 10^{-52}$; T5: 0.98 → 0.66 → 0.14, $\chi^2 = 211$, $p < 10^{-46}$), while page and layout files maintain stable high compliance throughout. T3 component files show a U-shape pattern instead (0.90 → 0.55 → 0.84).

**Sensitivity: AST-traversal function index.** Re-fitting the same GLMM with the AST-traversal function_index as predictor (sorting files alphabetically, then walking each file's source tree in declaration order) yields a steeper pooled slope (OR = 0.909, 95% Wald CI [0.901, 0.916]; $-0.0957$ log-odds per position) but the same qualitative shape: pooled attenuation is real and highly significant, the per-task heterogeneity persists, and T3 remains not detectable. The AST-traversal predictor incorporates within-file structural position alongside generation order, so we report the chronological metric as primary because it more directly indexes the construct of practical interest; both metrics agree on the qualitative finding.

### 4.7 Bayesian robustness

A parallel set of Bayesian GLMMs fitted in bambi on PyMC (4 chains, 2,000 post-warmup draws, $\hat{R} \leq 1.000$) is consistent with the frequentist analyses on every research question. The within-session attenuation slope on generation position is centred at $-0.0578$ log-odds per generation step (95% credible interval $[-0.0655, -0.0501]$); essentially all posterior mass is below zero and the Savage-Dickey Bayes factor against zero is at the numerical floor (effectively infinite). In odds-ratio terms: OR = 0.944 with a 95% credible interval of $[0.937, 0.951]$, matching the frequentist Wald estimate to four decimal places. For size, $BF_{10}$ = 0.096 under the default autoscaled prior (strong affirmative null on the Jeffreys scale; the posterior is roughly ten times more concentrated near zero than the prior). For conflict in the A2 pool, $BF_{10}$ = 0.053 (strong affirmative null). A prior-sensitivity sweep across Normal $(0, \sigma)$ priors on the target slope for $\sigma \in \{0.5, 1.0, 2.5, 5.0\}$ confirms the affirmative-null framing is robust to prior choice: $BF_{10}$ stays below 1/3 (the Jeffreys-scale "moderate evidence for null" threshold) for both effects under every prior tested (size $BF_{10}$ range $[0.019, 0.197]$; conflict $BF_{10}$ range $[0.023, 0.221]$). At the tightest prior tested ($\sigma = 0.5$) both effects are in the "moderate" range ($BF_{10}$ between 0.10 and 0.33); under wider priors both effects move into the "strong" or "extreme" ranges. Within the tested ranges, the size and conflict nulls are supported as affirmative null evidence rather than as failures to reject.

**Figure 4: Bayesian posterior densities.**

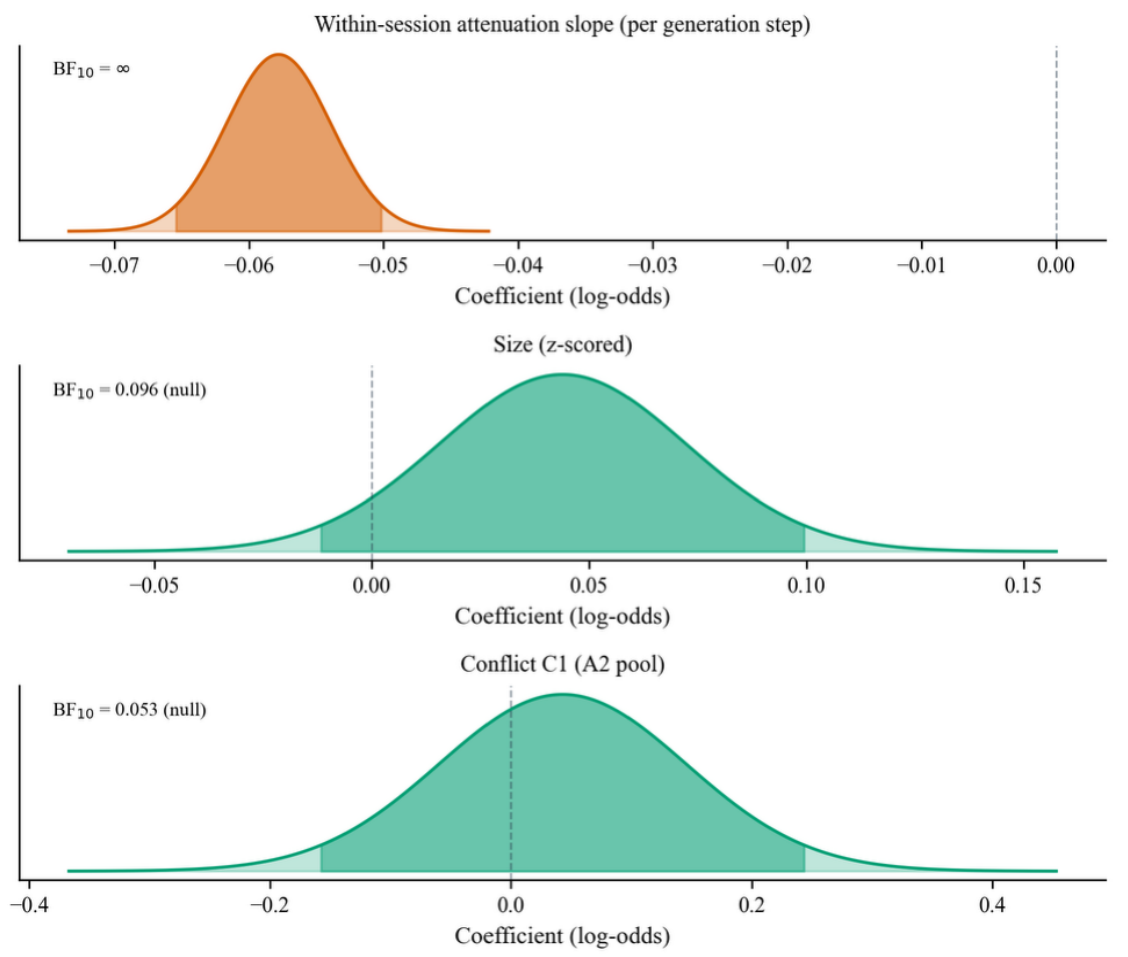

**Bayesian posterior densities for the three effects of theoretical interest, with 95% credible intervals shaded darker. Size and conflict are strong-affirmative nulls; the within-session attenuation slope is supported with effectively infinite Bayes factor. The posterior on size standardised by SD is centred just above zero with mass concentrated tightly around zero; the conflict posterior is similarly concentrated; the attenuation posterior is concentrated tightly around $-0.058$ log-odds per generation step.**

### 4.8 Follow-on investigations

We report four follow-on investigations briefly.

**Quality trade-off.** We measure task_completion_quality as the proportion of the task's test suite that passes on the agent's submitted code (0 to 1). The bivariate Spearman correlation between per-run ICR and task_completion_quality is weakly positive ($\rho = +0.104$, $p = 4.3 \times 10^{-4}$); after controlling for task identity, the ICR coefficient is not detectable (OR = 1.25, 95% CI [0.57,2.73], $p = 0.58$). **The // @tracked annotation has no detectable quality cost in our data.**

**First omission.** Among the 756 of 1,150 non-baseline runs (65.7%) that produced at least one non-compliant function, the median first-omission generation position was 4 (chronological metric; AST-traversal median is 2). Post-omission compliance is 55.0% across these runs.

**Figure 5: First-omission dynamics.**

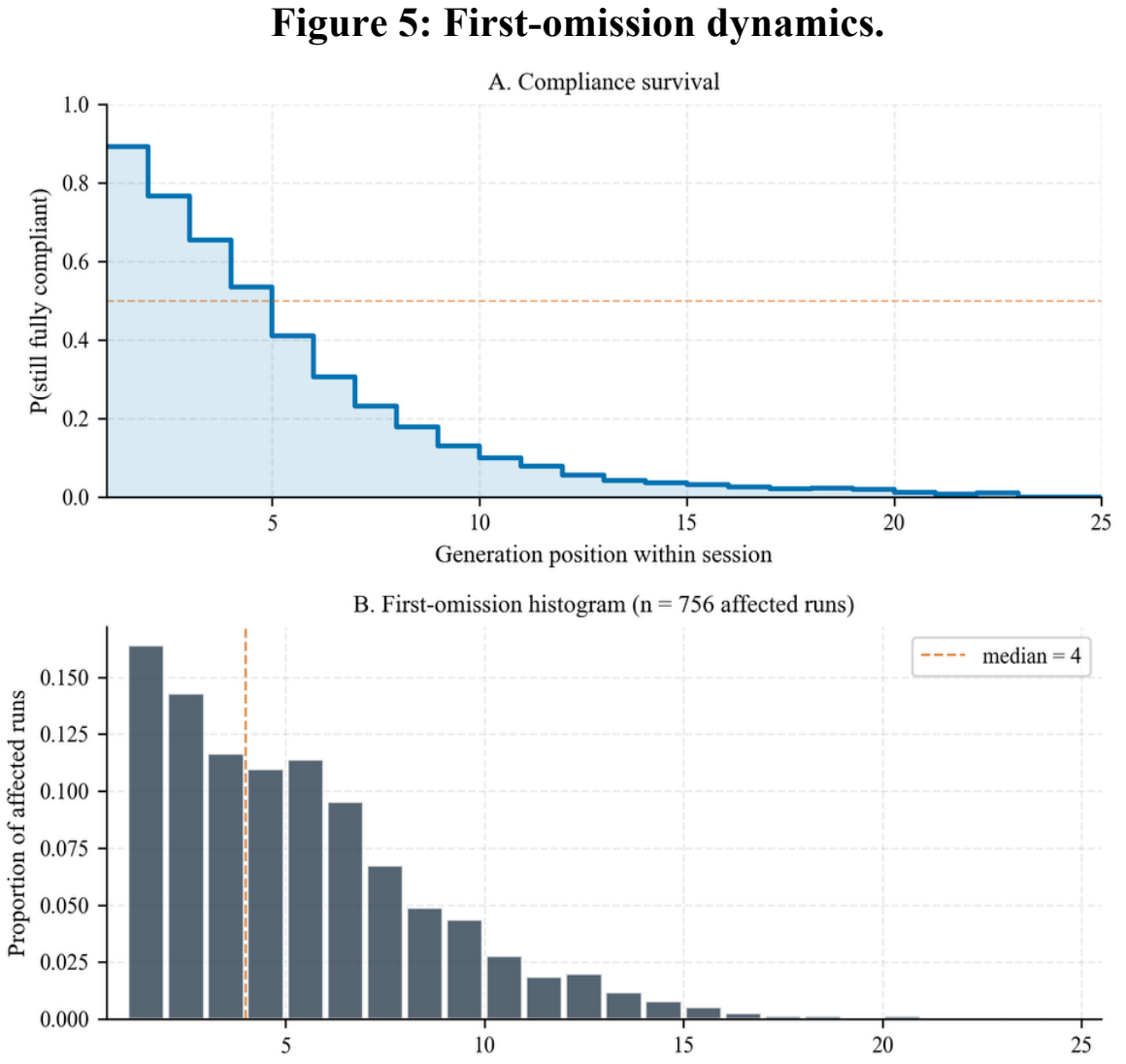

**First-omission dynamics on generation order. Panel (a): Kaplan-Meier-style survival of "still fully compliant", by generation position within a session; the median-survival position is marked with a dashed line. Panel (b): histogram of first-omission generation positions among the 756 affected runs; the distribution centres on a median of 4.**

**Acknowledgement.** The agent acknowledged the instruction explicitly in 112 of 1,150 runs (9.7%). Acknowledged functions had ICR 72.0% versus 62.8% for non-acknowledged (+9.2 pp, $\chi^2 = 19.04$, $p <$ $0.0001$, $|h| = 0.197$, small); 28% of acknowledged functions still did not comply; 63% of non-acknowledged complied silently.

**Session length.** Session-length proxies (turns, total cost, total functions) correlate with ICR but the correlations attenuate once task is included in a multiple regression; total functions per run is mechanically tied to the within-session attenuation above. No separate session-length finding.

## 5 Discussion

### 5.1 Within-session compliance attenuation

The largest effect we measured is within-session: each additional function the agent generates is associated with approximately 5.6% lower odds of complying with the target instruction on average (OR = 0.944), though the relationship is non-monotonic across the within-session range rather than a constant per-step effect. The within-session question was identified during analysis from descriptive inspection of per-run patterns rather than pre-specified as a research question, so we treat the finding as a follow-on investigation. The structural variables we manipulated do not produce contrasts of comparable magnitude; the within-session attenuation operates alongside any between-condition pattern in our data.

The per-function index is more closely associated with compliance than wall-clock time or turn count: the session-length / ICR association attenuates once total functions per run is included as a covariate, consistent with total function count being a more proximal predictor than session duration. The shape is also task-dependent: T5 is closest to a clean monotonic attenuation, T4 shows shape across all three polynomial orders, and T3 has a non-monotonic shape that averages to flat (Results §“Sensitivity: higher-order chronological shape”). Together these suggest the within-session signal depends on factors beyond session length alone, including task content.

The within-session attenuation is a function-level association. Mechanism-level investigation requires interventional studies we did not run, and we treat the mechanism question as open.

A separate compositional question is whether the slope reflects a single attenuation process or two regimes superimposed. Among the 65.7% of non-baseline runs that produced any non-compliance, the median chronological first-omission position was 4 (Results §“First omission”), so a substantial share of the attenuation mass sits in the first few generated functions, not later in the session. Post-first-omission compliance is 55.0% (not zero), so these runs are not simply giving up on the instruction; they show early non-compliance and then re-comply intermittently. The pooled OR = 0.944 averages across both regimes and does not separate them. The practitioner implications differ: early non-compliance suggests prompt-construction or configuration-loading failures (the agent does not pick up the instruction at session start); within-session drift suggests a mechanism that gradually weakens adherence to a successfully-loaded instruction over the course of generation. Disentangling them is a directed follow-up rather than a present finding.

A descriptive recovery in compliance at high generation positions (visible in Figure 1) is largely compositional in our data: component files in T4 and T5 continue to regress monotonically across position buckets, while page and layout files maintain stable high compliance throughout (Results §“Compositional shift in the late-session recovery”).

### 5.2 The size and conflict nulls are affirmative, not absent

Frequentist non-rejection is not the same as evidence for the null, so the Bayesian companion is the load-bearing test for the size and conflict findings. Both effects produce affirmative-null Bayes factors on the Jeffreys scale. Within the ranges we tested, a 25-line CLAUDE.md and a 500-line CLAUDE.md are indistinguishable across the sizes and positions we tested, and a secondary file containing a contradictory instruction does not measurably override the primary file’s instruction.

These nulls are useful for practitioners. Two common worries in the practitioner discourse, that configuration files grow too long and that AGENTS.md will undercut CLAUDE.md, do not produce detectable compliance penalties at the sizes and conflict format we tested. The size finding sits in tension with practitioner advice that emphasises keeping configuration compact [6]; the attention-dilution literature [11] is concerned with much longer context scales than we tested and is not strictly contradicted here.

### 5.3 No detectable position shape

Liu et al.’s “Lost in the Middle” [9] would predict a U-shape across the five positions we tested, with primacy and recency at the file top and bottom and a dip in the centre. We did not find one. Pairwise contrasts across the five positions all fail to survive multiple-testing correction, and an orthogonal- polynomial decomposition under the planned GLMM tests the shape claim directly: the quadratic component (which would capture a U-shape) is not detectable ($p = 0.94$), and neither linear, cubic, nor quartic components are detectable either ($p = 0.41, 0.39, 0.83$). The descriptive pattern is consistent with weak primacy and a small monotonic drift rather than a U-shape, but no shape component is detectable in our data within the levels we tested.

One tentative reading is that lost-in-the-middle is strongest for question-answering tasks with a single retrieval target, whereas an always-on configuration rule has no query cue to anchor recency on. The position null does not have Bayes-factor support, so the present results are equally compatible with a small underpowered U-shape we could not detect; this remains a hypothesis rather than a finding from the present design.

### 5.4 Task type as a single-contrast observation

The largest within-codebase contrast in our task set is between T4 (a refactor of an existing layout component) and T5 (a greenfield analytics-dashboard build), with similar average function counts per run (17.5 and 15.0 respectively). The 26.2-percentage-point gap is larger than the separation between any two structural conditions. The two tasks differ on many dimensions besides modify-vs-write (domain, helper-function ratios, file count touched, code patterns), all held constant within each task but varying across them. T4 is the only modify-existing task at the *task* level, so the modify-vs-write reading rests on a single between-task contrast; the Umami replication preserves the ranking but uses the same task design and does not independently vary the distinction.

We therefore do not endorse a "modifying is harder than writing" reading as a takeaway. The 26.2-percentage-point T4-versus-T5 gap is observed in our data but the dimensions on which T4 and T5 differ are multiple, and the present design does not isolate any one of them. Multi-task replication of modify-vs-write at matched complexity is required to test the framing; the present design supports the contrast but not the generalisation. Several within-task variables (function position within the session, file location, function size, pre-existing patterns in the file) plausibly shape per-function compliance; isolating their contributions requires a follow-up design that varies them deliberately. Whatever the underlying driver, task identity is a stronger predictor of compliance than any structural variable we manipulated, and T4 (the lowest-ICR task in our set) accounts for most of the contrast.

### 5.5 Cross-codebase: codebase factors shift the intercept

The Umami block adds three matched cells to the design, and the results from those cells should be interpreted as a one-codebase sanity check rather than as a structural-variable replication. Specifically: EV-02 (no configuration) replicates the zero baseline cleanly; EV-01 (matched A1/S3/P1/C0) gives one absolute ICR comparison against ixartz ME-03 and shows compliance approximately 11 percentage points lower on the larger codebase; EV-03 confounds A2 architecture with C1 conflict in a single cell, so the within-Umami EV-01 vs EV-03 contrast is not a clean architecture comparison and we do not treat it as independent confirmation of the ixartz architecture null. What the Umami data does support is that the absolute compliance level shifts down on the larger codebase while the zero baseline holds, consistent with codebase factors (size, framework, domain, code patterns) acting as an intercept shift rather than a slope or shape modifier.

This is useful external-validity context for the findings: the absolute number from any single codebase is not transportable. The structural-variable nulls reported in the previous sections are estimated on ixartz alone; their generalisation to other codebases, languages, and agents remains an open question.

### 5.6 Cross-model observations

The most informative cross-model contrast in our data is the matched-configuration comparison between Sonnet 4.6 and Opus 4.6. **Opus 4.6** ($n = 49$ valid runs at the matched OP-01 condition, $|h| = 0.264$) sits 12.8 percentage points below Sonnet 4.6 at identical configuration (S3 / P1 / A1 / C0, CLI 2.1.92). The within-session attenuation slope reproduces in direction on Opus 4.6 and the zero baseline replicates exactly, so the model difference is in the *intercept* of compliance rather than its dynamics: across the within-session range we tested, Opus 4.6 sits below Sonnet 4.6.

The standard practitioner prior is that more capable frontier models follow instructions better. Opus 4.6 was Anthropic's most capable model at the time the experiments were run, and the data go the other way at matched configuration on our task suite. We treat this as a single between-model contrast on a single task suite, codebase, and DV (a syntactically detectable marker), not a general claim that one model follows instructions better than the other: with only two CLI-matched models in the design we cannot separate model capability from model-specific instruction-tuning, and the ranking might well reverse on a different task suite or codebase that better suits Opus's training profile. The contrast does show that absolute compliance rates may not transport across models even within a single vendor's family at matched configuration; practitioners who calibrate on one model should not assume the same level on the next.

The observation is broadly compatible with the formal instruction-hierarchy literature [17], [18], where formal priorities have been shown to behave non-uniformly across models. The OP-01 versus OP-02 "reference plus stress" contrast is small and unstable in direction across the two Opus models (Opus 4.6: $+2.8$ pp; Opus 4.7: $-0.8$ pp; neither significant); the contrast varies three IVs simultaneously and is reported descriptively rather than as a structural-variable claim.

**Opus 4.7** is structurally uninterpretable in the present design. It produces no code on 18% of with-configuration trials and 40% of no-configuration baseline trials, with the non- production cells concentrated on two of the three multi-function tasks (T3 and T4; T5 is unaffected). On the trials where it does produce code, ICR is numerically higher than the other models'. We do not draw a model conclusion from this. Opus 4.7 ran on Claude Code CLI 2.1.123 (forced by an API schema change for adaptive thinking) while the other two models ran on CLI 2.1.92, and the 2.1.123 schema enables adaptive thinking by default. The Code-Production Rate gap and any cross-model comparison involving Opus 4.7 are jointly attributable to the model and to the CLI/thinking configuration; within this study they cannot be cleanly separated. Replication at matched CLI version is the appropriate follow-up.

### 5.7 Practical implications

The data point to three implications for practitioners writing or auditing a CLAUDE.md file, ranked by effect magnitude (Table 5).

The largest effect, within-session attenuation, is the practical centre of gravity: omissions occur during generation in our data, so in-session mechanisms are where the practical action sits.

**Table 5: Practical implications.**

| Finding | Practitioner action |
|---|---|
| Within-session attenuation (largest effect) | Compliance trends downward on average as the agent writes more code, with rate and shape varying by task. Layer in-session mechanisms (agent hooks that re-surface the rule, agent-managed memory) with post-hoc enforcement (linters, CI checks); the present design does not test multi-file configuration as a remediation, and the architecture-null result does not support it within the levels we tested. |
| Task-type asymmetry (second-largest) | For tasks like T4 (a multi-component refactor of an existing layout), where compliance is lowest in our data, prefer enforcement mechanisms that travel with the code: agent hooks that intercept tool calls during the work, linter rules, and CI checks. The single-contrast caveat in §"Task type as a single-contrast observation" cautions against generalising this to all modify-existing tasks. |
| File length 25 to 500 lines (null) | Focus on content rather than length within this range; length did not produce a detectable compliance contrast in our data. |

### 5.8 Bounds on the null findings

The Bayesian Bayes factors for size and conflict license a stronger claim than "failure to reject", but only within the levels we tested. A 5,000-line CLAUDE.md exists in some large repositories and is well outside our design; the dilution that theory predicts at extreme scales may well appear there. A more salient conflict format, for instance an AGENTS.md instruction that lexically overlaps with the CLAUDE.md instruction rather than semantically opposing it, may produce a measurable effect that our format did not. The position and architecture nulls do not have Bayes-factor support and should be read as failures to reject rather than as affirmative nulls. The cross-model observations, the cross-codebase absolute drop, and the Opus 4.7 sentinel pattern are exploratory and should be replicated at larger samples and with the same CLI version across models before being relied upon.

## 6 Limitations

The dependent variable is AST-detected presence of a single static annotation (// @tracked). The scoring script is deterministic and version-controlled, so identical agent diffs always receive identical scores. The primary internal-validity risk is stochastic variation in the agent itself: Claude Code does not expose a sampling seed, so repeated runs of the same condition and task produce different outputs. We mitigate this with $n = 50$ runs per condition and function-level analysis that aggregates across both runs and functions. A secondary risk is configuration-injection bleed: the test repository is reset to its pinned baseline commit between every run, but cache locations the CLI may touch (model context caches, temporary files in ~/.claude/) were not exhaustively audited. The zero-baseline replicates exactly across the four independent baseline cells, consistent with no contamination of the compliance signal.

ICR as we measure it is an annotation-insertion rate. We treat it as an easily-detectable proxy for instruction adherence; whether and how it transfers to more demanding instructions (e.g. "always run the test suite before reporting completion” or “never modify the schema without updating the migration”) is an open question. The within-session attenuation we measured may transfer in shape; the absolute ICR values are not portable to other markers. The specific phrasing of the target instruction (“MUST begin with the comment // @tracked”) may be more or less effective than a semantically equivalent alternative; we did not vary instruction phrasing, and the practitioner literature [6], [18] reports that phrasing effects can be substantial.

The study covers two TypeScript codebases (the ixartz Next.js boilerplate and Umami), three Anthropic frontier models (Sonnet 4.6, Opus 4.6, Opus 4.7), and the Claude Code CLI in single-turn --print mode. Other languages, other coding agents (Codex, Cursor, Gemini CLI, opencode), and other configuration formats (AGENTS.md without CLAUDE.md, repository-level READMEs) are not covered and may behave differently. We studied only repository-level static configuration files (CLAUDE.md, AGENTS.md, and nested CLAUDE.md files), not agent-managed memory files that the agent writes and re-reads on later turns through different mechanisms. The cross-codebase comparison on Umami documents an 11-percentage-point absolute-ICR drop relative to ixartz at matched configuration, consistent with codebase factors shifting absolute compliance; the magnitude in any individual case is not estimable from one comparison. The Sonnet-Opus 4.6 gap of 12.8 percentage points is a reminder that model-specific effects are not small even within a single vendor’s family.

The five coding tasks (T1–T5; see Methodology Table 3) cover feature-addition work on a Next.js boilerplate but are not a representative sample of the broader coding-task space. Tasks like multi-file refactoring, debugging, infrastructure work, dependency upgrades, and code review are not represented in our design; both the absolute compliance levels we report and the per-task within-session attenuation shapes may differ on other task types. The 26.2-percentage-point T4-versus-T5 gap in particular is a single between-task observation and does not generalise to a modify-vs-write claim across the full task space.

The single-turn --print harness is a meaningful constraint on external validity. Real Claude Code usage is multi-turn: the agent reads the configuration file at session start and then operates across many turns, with various compaction and summarisation behaviours that may re-surface or evict the configuration mid- session. Single-turn --print mode is a cleaner setting for controlled measurement but a constrained subset of how configuration files are actually used in production. The within-session attenuation we observe is a function-level association estimated under single-turn execution; under multi-turn execution the dynamics may differ depending on whether the configuration is re-injected, compacted, or summarised between turns. We do not claim how compliance behaves under agents that aggressively compact or summarise context mid-session, or outside the Claude Code CLI. The position null also has a context-length boundary: our position sweep is at S3 = 250 lines, which corresponds to roughly two to three thousand tokens of CLAUDE.md content. Liu et al.’s “Lost in the Middle” U-shape is canonically observed at much longer contexts, so the absence of a U-shape here cannot be read as an absence of the U-shape at the longer-context regime where that literature was developed. A position sweep at the longer file length (e.g. our S4 = 500 lines, or larger) would be the appropriate follow-up.

Opus 4.7 is confounded along three dimensions, not two. Sonnet 4.6 and Opus 4.6 ran on Claude Code CLI 2.1.92; Opus 4.7 ran on CLI 2.1.123 because an API schema change for adaptive thinking forced an upgrade. The 2.1.123 schema enables adaptive thinking by default, while CLI 2.1.92 does not. So the Opus 4.7 condition differs from the Sonnet/Opus 4.6 conditions in three respects simultaneously: (i) model version, (ii) CLI version, and (iii) adaptive-thinking mode. The elevated clarification-before-code rate and the higher conditional ICR on Opus 4.7 cannot be attributed to any one of these three; in particular, what reads as a model-level finding could equally be a finding about adaptive thinking. A model × position interaction GLMM that includes Opus 4.7 in the chronological pool returns a detectable interaction ($\chi^2 = 16.72, df = 2, p = 0.0002$), but this is not informative because the per-cell decomposition shows the detection is driven by Opus 4.7’s steeper conditional slope; the CLI-matched-only equivalent (Sonnet 4.6 + Opus 4.6 + Umami Sonnet 4.6) returns $\chi^2 = 2.23, df = 2, p = 0.327$ (see Results §“Cross-codebase consistency”). We therefore report the Opus 4.7 numbers for completeness and flag the underlying confound with every cross-model claim; we do not draw a model-only conclusion from them.

The within-session pattern was identified during analysis, and the structural variables we manipulated do not speak to its mechanism. A covariate-adjusted check on function-type composition (size, within-file rank, single-line share) showed composition shifts substantially with position, but observational data cannot distinguish a confound from a downstream symptom of the same mechanism. The appropriate follow-up is an experiment that manipulates composition independently of session position.

The within-session attenuation is estimated within sessions of up to 81 generated functions (median 18). Whether the per-step shape persists, plateaus, or recovers at the much longer session lengths used in production multi-turn agent sessions is not estimable from our design; we do not claim a mechanism that would extrapolate beyond our tested range. Each of our 1,650 sessions is independent, with no carry-over between runs and a zero baseline that replicates exactly across baseline cells, so the attenuation is a within-session phenomenon rather than a long-term degradation across runs.

The primary ixartz blocks at $n = 50$ per condition produce roughly 500 function-level observations per cell and are adequately powered for effects above approximately 5 to 6 percentage points at the pairwise-contrast level (see Results §"Structural variables show no detectable effect" for the Bayesian and frequentist evidence classes); per-IV power for the joint GLMM factor-LRT varies by degrees of freedom and is reported in Methodology §"Sample size". The cross-model Opus blocks were designed as a "reference plus extreme" pair rather than a within-Opus replication, so the OP-01 versus OP-02 contrast does not isolate any single structural variable on Opus. The within-session attenuation GLMM exhibits complete separation on T1 and T2 (1 to 2 functions per run), so the attenuation-by-task interaction is fitted on T3, T4 and T5 only. The Umami EV-03 condition combines architecture (A2) with conflict (C1) and cannot isolate either factor. Our analyses use Benjamini-Hochberg correction for post-hoc pairwise tests and no correction for the planned GLMM contrasts; a stricter Bonferroni correction at $\alpha/7 = 0.0071$ for the seven structural contrasts in the design also yields no surviving contrasts, and the within-session attenuation clears any reasonable multiple-testing threshold by many orders of magnitude.

## 7 Conclusion

This study investigated whether four features of agent configuration files (size, instruction position, file architecture, and inter-file conflict) affect the rate at which a frontier coding agent follows a well-placed target instruction. Across 1,650 controlled sessions and 16,050 function-level observations on two TypeScript codebases and three frontier models, we find that none of the four manipulated structural variables produces a detectable contrast within the tested ranges and none of the three two-way interactions reaches significance after correction. The size and conflict nulls are supported by Bayes factors of approximately 0.05 to 0.10 (the data favour the null over a unit-effect alternative by a factor of approximately ten to twenty on the Jeffreys scale, within the tested ranges); the position and architecture nulls are failures to reject without Bayes-factor support, and the design rules out structural effects larger than approximately 6 percentage points on the largest pairwise contrasts but cannot rule out smaller effects.

A follow-on analysis, identified during data inspection rather than set out as a research question in advance, found the largest effect we measured: each additional function the agent generates within a session is associated with approximately 5.6% lower odds of compliance per generation step (OR = 0.944), though the relationship is non-monotonic rather than a constant per-step effect, and per-task variation is substantial (this effect is detectable on two of the three multi-function coding tasks tested).

The largest within-codebase contrast we measure is a 26.2 percentage-point gap between two specific tasks (a refactor of an existing layout component and a greenfield analytics- dashboard build, with similar average function counts per run). We report this as a single-contrast observation between two specific tasks rather than a modify-vs-write generalisation.

For practitioners writing or auditing a CLAUDE.md file, four practical implications follow within the tested ranges. **Expect compliance to trend downward as the agent writes more code within a session**, with rate and shape varying by task; instructions that must hold across long sessions benefit from agent hooks that re-surface the rule mid-session (PreToolUse, PostToolUse, Stop), agent-managed memory or context-engineering patterns, and post-hoc enforcement by tooling (linters, CI checks). The architecture null in our data does not support distributing the configuration across multiple files as a remediation within the levels we tested. **Compliance varies systematically with task type in our data**: within the conditions we tested, the structural variables we manipulated did not produce detectable contrasts, while task type produced larger contrasts than any of them, so practitioners should treat instructions as task-type-conditional rather than file-layout-conditional. **Within 25 to 500 lines, file length did not produce a detectable compliance contrast in our data**, and within the format we tested, **a contradicting instruction in a secondary file did not detectably reduce compliance** with the primary file.

Three directions warrant follow-up. First, the within-session attenuation should be examined on other agent harnesses (Cursor, Codex, opencode, Gemini CLI), other model families (GPT, Gemini, Kimi), and other languages (Python, Rust, Go) to determine how far the per-function shape generalises. Second, our target instruction is deliberately trivial; the rate at which more demanding instructions (multi-step procedural rules, conditional rules, instructions requiring cross-file reasoning) decline within a session is likely to differ, and a natural extension varies instruction complexity alongside the structural variables. Third, the cross-model observations are confounded with CLI version and warrant follow-up at larger sample sizes with the same CLI across models. Within the conditions we tested, the structural variables in our design were not the largest sources of compliance variation we measured; the within-session attenuation and task identity were.

## Appendix A: Full Condition Matrix

Table A1 lists every experimental condition across the primary (ixartz / Sonnet 4.6), ecological (Umami / Sonnet 4.6), and two cross-model datasets (ixartz / Opus 4.6 and ixartz / Opus 4.7). Each ixartz Sonnet, Umami, Opus 4.6 and Opus 4.7 condition was executed with 50 runs. Conditions marked ★ are shared reference cells (ME-03 is the main reference used across the main-effects block and the clean-configuration dataset comparisons). C1 conditions necessarily use A2 or A3 architecture because the conflict instruction lives in a secondary file. Sonnet 4.6 and Opus 4.6 ran on Claude Code CLI 2.1.92; Opus 4.7 ran on CLI 2.1.123 (forced by an API schema change for adaptive thinking).

**Table A1: Full condition matrix across the four datasets.**

| Block | ID | Size | Position | Architecture | Conflict | Isolated factor |
|---|---|---|---|---|---|---|
| Main effects | ME-01 | S1 | P1 | A1 | C0 | Size (min) |
| Main effects | ME-02 | S2 | P1 | A1 | C0 | Size |
| Main effects | ME-03 ★ | S3 | P1 | A1 | C0 | *reference* |
| Main effects | ME-04 | S4 | P1 | A1 | C0 | Size (max) |
| Main effects | ME-05 | S3 | P2 | A1 | C0 | Position |
| Main effects | ME-06 | S3 | P3 | A1 | C0 | Position (middle) |
| Main effects | ME-07 | S3 | P4 | A1 | C0 | Position |
| Main effects | ME-08 | S3 | P5 | A1 | C0 | Position (bottom) |
| Main effects | ME-09 | S3 | P1 | A2 | C0 | Architecture |
| Main effects | ME-10 | S3 | P1 | A3 | C0 | Architecture (nested) |
| Main effects | ME-11 | S3 | P1 | A2 | C1 | Conflict |
| Size × Position | SP-01 | S2 | P3 | A1 | C0 | Size × Position |
| Size × Position | SP-02 | S2 | P5 | A1 | C0 | Size × Position |
| Size × Position | SP-03 | S4 | P3 | A1 | C0 | Size × Position |
| Size × Position | SP-04 | S4 | P5 | A1 | C0 | Size × Position |
| Size × Arch. | SA-01 | S2 | P1 | A2 | C0 | Size × Architecture |
| Size × Arch. | SA-02 | S2 | P1 | A3 | C0 | Size × Architecture |
| Size × Arch. | SA-03 | S4 | P1 | A2 | C0 | Size × Architecture |
| Size × Arch. | SA-04 | S4 | P1 | A3 | C0 | Size × Architecture |
| Conflict × Position | CP-01 | S3 | P3 | A2 | C1 | Conflict × Position |
| Conflict × Position | CP-02 | S3 | P5 | A2 | C1 | Conflict × Position |
| Conflict × Position | CP-03 | S3 | P3 | A2 | C0 | Conflict × Position |
| Conflict × Position | CP-04 | S3 | P5 | A2 | C0 | Conflict × Position |
| Baseline | BL-01 | n/a | n/a | none | n/a | No configuration |
| Ecological (Umami) | EV-01 | S3 | P1 | A1 | C0 | Clean config / larger codebase |
| Ecological (Umami) | EV-02 | n/a | n/a | none | n/a | Baseline on larger codebase |
| Ecological (Umami) | EV-03 | S3 | P1 | A2 | C1 | Split + conflict on larger codebase |
| Opus 4.6 (CLI 2.1.92) | OP-01 | S3 | P1 | A1 | C0 | Clean config / Opus 4.6 |
| Opus 4.6 (CLI 2.1.92) | OP-02 | S3 | P3 | A2 | C1 | Worst case (split + middle + conflict) / Opus 4.6 |
| Opus 4.6 (CLI 2.1.92) | OP-03 | n/a | n/a | none | n/a | Baseline / Opus 4.6 |
| Opus 4.7 (CLI 2.1.123) | OP-01 | S3 | P1 | A1 | C0 | Clean config / Opus 4.7 |
| Opus 4.7 (CLI 2.1.123) | OP-02 | S3 | P3 | A2 | C1 | Worst case (split + middle + conflict) / Opus 4.7 |
| Opus 4.7 (CLI 2.1.123) | OP-03 | n/a | n/a | none | n/a | Baseline / Opus 4.7 |